\documentclass[revtex4]{emulateapj}

\usepackage{lscape}
\usepackage{graphicx}
\newcommand{\mytilde}{\raise.17ex\hbox{$\scriptstyle\sim$}}
\newcommand{\brk}[1]{$[$#1$]$}
\newcommand{\kepler}{{\it Kepler}}

\newcommand{\Msun}{${\rm M_\odot}$}
\newcommand{\Rsun}{${\rm R_\odot}$}

\newcommand{\Rear}{${\rm R_\oplus}$}

\newcommand{\Mstar}{${\rm M_\star}$}
\newcommand{\Rstar}{${\rm R_\star}$}
\newcommand{\Porb}{${\rm P_{orb}}$}
\newcommand{\Teff}{${\rm T_{\rm eff}}$}
\newcommand{\Teq}{${\rm T_{eq}}$}

\newcommand{\Rplanet}{${\rm R_P}$}
\newcommand{\Logg}{$\log{\rm g}$}
\newcommand{\Feh}{${\rm [Fe/H]}$}
\newcommand{\Tdur}{${\rm T_{dur}}$}
\newcommand{\NewPC}{472}
\newcommand{\TotPC}{2,738}
\newcommand{\TotTargets}{2,017}

\begin{document}

\title{Planetary Candidates Observed by \kepler\ IV: Planet Sample From Q1-Q8 (22 Months)}

\author{
Christopher~J.~Burke$^{1}$,
Stephen~T.~Bryson$^{2}$,
F.~Mullally$^{1}$,
Jason~F.~Rowe$^{1}$,
Jessie~L.~Christiansen$^{3}$,
Susan~E.~Thompson$^{1}$,
Jeffrey~L.~Coughlin$^{1}$,
Michael~R.~Haas$^{2}$,
Natalie~M.~Batalha$^{2}$,
Douglas~A.~Caldwell$^{1}$,
Jon~M.~Jenkins$^{1}$,
Martin~Still$^{4}$,
Thomas~Barclay$^{4}$,
William~J.~Borucki$^{2}$,
William~J.~Chaplin$^{5}$$^{,6}$,
David~R.~Ciardi$^{3}$,
Bruce~D.~Clarke$^{7}$,
William~D.~Cochran$^{8}$,
Brice-Olivier~Demory$^{9}$,
Gilbert~A.~Esquerdo$^{10}$,
Thomas~N.~Gautier III$^{7}$,
Ronald~L.~Gilliland$^{11}$,
Forrest~R.~Girouard$^{12}$,
Mathieu~Havel$^{2}$,
Christopher~E.~Henze$^{2}$,
Steve~B.~Howell$^{2}$,
Daniel~Huber$^{2}$,
David~W.~Latham$^{10}$,
Jie~Li$^{1}$,
Robert~C.~Morehead$^{13}$$^{,21}$,
Timothy~D.~Morton$^{14}$,
Joshua~Pepper$^{15}$$^{,16}$,
Elisa~Quintana$^{1}$,
Darin~Ragozzine$^{17}$,
Shawn~E.~Seader$^{1}$,
Yash~Shah$^{18}$,
Avi~Shporer$^{19}$,
Peter~Tenenbaum$^{1}$,
Joseph~D.~Twicken$^{1}$,
Angie~Wolfgang$^{20}$$^{,21}$
}
\affil{$^{1}$SETI Institute/NASA Ames Research Center, Moffett Field, CA 94035}
\affil{$^{2}$NASA Ames Research Center, Moffett Field, CA 94035}
\affil{$^{3}$NASA Exoplanet Science Institute/Caltech, Pasadena, CA 91125}
\affil{$^{4}$Bay Area Environmental Research Institute/NASA Ames Research Center, Moffett Field, CA 94035}
\affil{$^{5}$School of Physics and Astronomy, University of Birmingham, Edgbaston, Birmingham, B15 2TT, UK }
\affil{$^{6}$Stellar Astrophysics Centre (SAC), Department of Physics and Astronomy, Aarhus University, Ny Munkegade 120, DK-8000 Aarhus C, Denmark}
\affil{$^{7}$Jet Propulsion Laboratory/California Institute of Technology, Pasadena, CA 91109}
\affil{$^{8}$Astronomy Department and McDonald Observatory, The University of Texas, Austin TX 78712}
\affil{$^{9}$Department of Earth, Atmospheric and Planetary Sciences, Massachusetts Institute of Technology, 77 Massachusetts Ave., Cambridge, MA 02139}
\affil{$^{10}$Harvard-Smithsonian Center for Astrophysics, 60 Garden Street, Cambridge, MA 02138}
\affil{$^{11}$Center for Exoplanets and Habitable Worlds, The Pennsylvania State University, University Park, PA 16802}
\affil{$^{12}$Orbital Sciences Corporation/NASA Ames Research Center, Moffett Field, CA 94035}
\affil{$^{13}$Department of Astronomy and Astrophysics, The Pennsylvania State University, University Park, PA 16802}
\affil{$^{14}$Department of Astrophysics, California Institute of Technology, MC 249-17, Pasadena, CA 91125}
\affil{$^{15}$Department of Physics, Lehigh University, 16 Memorial Drive East, Bethlehem, PA 18015}
\affil{$^{16}$Department of Physics and Astronomy, Vanderbilt University, 6301 Stevenson Center, Nashville, TN 37235}
\affil{$^{17}$Department of Physics and Space Sciences, Florida Institute of Technology, 150 West University Blvd., Melbourne, FL 32901}
\affil{$^{18}$University of California, Berkeley, Berkeley, CA 94720}
\affil{$^{19}$Division of Geological and Planetary Sciences, California Institute of Technology, Pasadena, CA 91125}
\affil{$^{20}$Department of Astronomy and Astrophysics, University of California, Santa Cruz, CA 95064}
\affil{$^{21}$NSF Graduate Research Fellow}

\email{christopher.j.burke@nasa.gov}

\begin{abstract}

We provide updates to the \kepler\ planet candidate sample based upon
nearly two years of high-precision photometry (i.e., Q1-Q8).  From an
initial list of nearly 13,400 Threshold Crossing Events (TCEs), 480
new host stars are identified from their flux time series as
consistent with hosting transiting planets.  Potential transit signals
are subjected to further analysis using the pixel-level data, which
allows background eclipsing binaries to be identified through small
image position shifts during transit.  We also re-evaluate
\kepler\ Objects of Interest (KOI) 1-1609, which were identified early
in the mission, using substantially more data to test for background
false positives and to find additional multiple systems.  Combining
the new and previous KOI samples, we provide updated parameters for
\TotPC\ \kepler\ planet candidates distributed across
\TotTargets\ host stars.  From the combined \kepler\ planet
candidates, \NewPC\ are new from the Q1-Q8 data examined in this
study.  The new \kepler\ planet candidates represent $\sim$40\% of the
sample with \Rplanet$\sim$1\Rear\ and represent $\sim$40\% of the low
equilibrium temperature (\Teq$<$300~K) sample.  We review the known
biases in the current sample of \kepler\ planet candidates relevant to
evaluating planet population statistics with the current
\kepler\ planet candidate sample.

\end{abstract}


\section{Introduction}\label{sec:intro}

The NASA \kepler\ spacecraft delivers high precision photometric
observations to identify large samples of transiting planets around
stars in the Milky Way Galaxy.  One of its primary science drivers is
to extend our knowledge of extrasolar planets to the regime of
Earth-size planets orbiting stars like the Sun \citep{BOR10}.  The
\kepler\ project has released a series of papers incrementally
increasing the planet candidate discoveries identified with \kepler\
data \citep{BOR11A,BOR11B,BAT13}.  This study is the continuation of
this series applied to transiting planet signals detected in 8
quarters, Q1-Q8, (nearly two years) worth of \kepler\ data.  In addition
to the new planet candidates, we re-evaluate the \kepler\ planet
candidates from \citet{BOR11A,BOR11B} that were announced using the
earliest available \kepler\ data.  Re-evaluating the earliest \kepler\
planet candidates increases the baseline of observations and takes
advantage of the more refined techniques for evaluating the
reliability and source of the transit signal.

The \kepler\ planet candidate sample is the basis for a wide variety
of exoplanetary studies and discoveries.  A subset of the
\kepler\ planet candidate sample has been confirmed using radial
velocity follow up
\citep[e.g.,][]{DUN10,LAT10,BAT11,END11,SAN11,HEB13}, statistical
analysis of the \kepler\ flux time series in order to rule out stellar
binary signals \citep[e.g.,][]{TOR11,MOR11,FRE11,BOR12,WAN13,BAR13},
and transit time variations
\citep[e.g.,][]{HOL10,LIS11A,FAB12A,FOR12,STE12,XIE13}.  Studying the
population of multiple planet candidate systems provides insight into
the formation, migration, and dynamical interaction processes that
result in the planets observed by
\kepler\ \citep[e.g.,][]{FOR11,LIS11B,FAB12B,REI12,HAN13,BATY13}.  The
underlying planet population in the Galaxy can be determined from the
observed planet candidate sample from \kepler\ using a thorough
understanding of the selection effects and sources of contamination
\citep[e.g.,][]{YOU11,HOW12,FRE13,CHR13,DRE13}.

In \S~\ref{sec:data} the \kepler\ data analysis pipeline is reviewed and
the process used to select the transit signals for analysis is
described.  \S~\ref{sec:triage} \& \S~\ref{sec:vetting} detail the
analysis techniques employed using \kepler\ data alone to ensure a high
degree of probability that the transit signal originates from the
target star under observation and eliminate possible sources of false
positives.  The purpose of this analysis is to classify the transit
signals as either a planet candidate or false positive.  The transit
signals are fit to a planet model, \S~\ref{sec:parameters}, to
determine planet parameters after assigning stellar parameters
following the procedure outlined in \citet{BAT13}.  We describe the
resulting population of planet candidates discovered by \kepler\ in
\S~\ref{sec:results}, and we conclude the study in
\S~\ref{sec:conclusion}.

\section{Transit Signal Detection}\label{sec:data}

Identification of planet candidates in \kepler\ data begins with
output from the \kepler\ science pipeline.  The \kepler\ pipeline
converts the raw instrument output of the \kepler\ spacecraft into a
format usable by the scientific community \citep[see][for an
  overview]{JEN10}.  Here we summarize only the Transiting Planet
Search (TPS) module of the \kepler\ pipeline as it performs the
transit signal detection using output of the earlier modules of the
\kepler\ pipeline that provide instrument corrected aperture flux time
series data.  TPS empirically determines the noise in the flux time
series \citep[combined differential photometric precision,
  CDPP,][]{CHR12} of each target to search for potential planet
candidates (threshold crossing events, TCEs) \citep{JEN02,TEN12}.  For
a transit signal in \kepler\ data to be defined as a TCE, the combined
signal-to-noise ratio (SNR) from multiple transit events (the multiple
event significance, MES) must be above a preset threshold, MES$>$7.1
\citep{JEN02}.  In addition, a transit signal must have a ratio of the
MES to the strongest single transit event SNR greater than $\sqrt{2}$
in order to qualify as a TCE \citep{TEN12}.  The criteria for
identifying a TCE has evolved through time, and the specifics
described above pertain to what was used in the Q1-Q8
\kepler\ pipeline run.  The input light curve input to TPS for the
Q1-Q8 pipeline run was generated by the Pre-search Data Conditioning
(PDC) algorithm as described by \citet{TWI10A}.

In its simplest form a TCE represents a transit candidate by
specifying the target \kepler\ Input Catalog (KIC) identifier
\citep{BRO11}, ephemeris period, ephemeris epoch, transit duration
estimate, and transit depth estimate.  In November 2011, the Q1-Q8 (22
months of data) pipeline run generated one or more TCEs for
$\sim$13,400 \kepler\ targets out of $\sim$191,000 targets searched.

\section{TCE Triage}\label{sec:triage}

The majority of TCEs ($\gtrsim$80\%) are not valid planet candidates.
Contributing are numerous types of astrophysical variability: stellar
oscillations \citep{AER10}, overcontact binaries \citep{SIR03}, tidal
dynamic distortions \citep{THO12}, and broad-band ``red noise''.  In
addition, the TCE population is contaminated by signals due to
instrumental effects: thermal transients, pixel sensitivity dropouts,
pattern noise, and video crosstalk \citep{CAL10,KOL10,STU12}.  Most of
these contaminants do not produce the archetypical signal for a
transiting planet, characterized by repetitive, isolated,
limb-darkened events with out-of-transit noise that averages down as
expected for independent data.

In order to efficiently remove the contaminating signals in the TCE
sample, all TCEs undergo a visual inspection of the \kepler\ light
curve data phase folded on the TCE ephemeris.  A standardized data
plot is generated for the TCE found for a given target in TPS.  The
data plot employs the aperture flux time series generated by the
Photometric Analysis (PA) module of the pipeline \citep{TWI10B}.  For
plotting, the flux time series is median detrended with a moving
window of 2 day duration and the resulting relative flux time series
is phase folded on the TCE ephemeris.  We perform this ``triage''
stage for the $\sim$13,400 TCEs, using the phase folded relative flux
time series.  The TCE is either accepted as a potential planet
candidate from visual inspection of the phase folded flux time series
and moves onto the next stage of vetting, or the TCE is eliminated
from further consideration as a planet candidate.

During ``triage'', 565 TCEs were identified around 480 new
\kepler\ targets that did not have previously known transit signal
detections.  The TCEs that pass the visual inspection ``triage'' stage
are designated as \kepler\ Objects of Interest (KOI).  A TCE that
passes ``triage'' is assigned a KOI number in order to catalog the
detection and to move forward in its analysis as a potential planet
candidate.  The Q1-Q8 TCEs that passed ``triage'' were assigned KOI
numbers in the range 2668$\leq$KOI$\leq$3149.  It is important to
emphasize that a KOI at this earliest stage has not been vetted
against the full complement of \kepler\ data and analysis tests as
described in \S~\ref{sec:vetting}.  The KOI sample before
dispositioning still has a high proportion of false positives due to
stellar binaries, instrumental artifacts, and other astrophysical
variability.  In subsequent dispositioning (described in
\S~\ref{sec:vetting}) $\sim$40\% of the new KOIs were given a false
positive designation.

\section{\kepler\ Object of Interest Dispositioning}\label{sec:vetting}

For a newly created KOI to be dispositioned as a planet candidate it
must pass further scrutiny using \kepler\ data.  Primarily, the KOI
dispositioning examines both the flux time series data, for
consistency with the expectation of a transiting planet signal, and
the pixel-level time series data, for consistency with the expectation
that the signal originates from the target of interest in the
aperture.  The dispositioning process follows the general procedure
outlined in \citet{BAT13}.  The present updates to the KOI sample
result from dispositioning two groups of \kepler\ targets.  The first
group of targets are new KOIs that were identified in the Q1-Q8 data
as outlined in \S~\ref{sec:data}~\&~\ref{sec:triage}.  The new KOIs in
this group are in the range 2668$\leq$KOI$\leq$3149.  The second group
of targets are the earliest KOIs (KOI number $\leq$1609) from the
first two \kepler\ planet candidate catalogs \citep{BOR11A,BOR11B}.
These KOIs are re-evaluated to take advantage of the substantially
increased data baseline and a more uniform set of dispositioning
criteria and procedures.  Overall, between the two groups of KOIs, we
evaluated $\sim$1900 KOIs around $\sim$1500 targets.  The most current
analysis of KOIs intermediate between these two groups,
1610$\leq$KOI$\leq$2667, is published in \citet{BAT13}.  The KOI
sample of \citet{BAT13} also added new KOIs in the $\leq$ 1609 range
that were discovered in multiplanet systems that we do not revisit in
this study.  As described in \S~\ref{sec:results},
Table~\ref{tab:parm} contains a binary flag to indicate whether the
KOI was vetted during this study.

Although the new KOIs reported here were discovered from the search of
Q1-Q8 data, the actual data products used for dispositioning were from
a Q1-Q10 (28 months) pipeline run.  Results from the Q1-Q10 pipeline
run were the most current pipeline products at the disposal of the
authors at the outset of the dispositioning (May 2012).  For the two
groups of KOIs that are a part of this study, we also dispositioned
any new TCEs that were discovered in the Q1-Q10 data and included them
in this KOI sample.  We did not evaluate new TCEs from the Q1-Q10
pipeline run for the intermediate, 1610$\leq$KOI$\leq$2667, targets or
other Q1-Q10 targets that were not previously known to have KOIs.
This inhomogeneity in the planet candidate sample has implications for
statistical studies of the underlying planet population as discussed
in \S~\ref{sec:complete}.

KOI dispositioning is primarily based upon a report and its separate
summary that are generated in the Data Validation (DV) module of the
pipeline \citep{WU10}.  DV report summaries are available from the
subsequent Q1-Q12 pipeline run hosted at the NASA Exoplanet
Archive\footnote{http://exoplanetarchive.ipac.caltech.edu}.  The
Q1-Q12 results are not available for all the KOIs examined in this
study since the pipeline runs are independent and subsequent runs are
not guaranteed to identify the same transiting signal.  95\% of the
cumulative \kepler\ planet candidate sample have Q1-Q12 DV products
available at the NASA Exoplanet Archive.

The DV report summary provides the \kepler\ flux time series and
pixel-level data tests that are most relevant for dispositioning the
KOI into one of three categories: planet candidate, needs further
scrutiny, or false positive.  The criteria and statistical tests for
dispositioning a KOI are outlined in \citet{BAT13} and are discussed
further in \S~\ref{sec:fluxvet} and \S~\ref{sec:centroidvet}.  Every
KOI had at least two individuals independently evaluate all criteria
in order to judge its planet candidate status.  KOIs with unanimous
evaluation as a planet candidate or false positive are dispositioned
as a planet candidate or false positive, respectively.  KOIs with a
discordant disposition are given additional scrutiny.  After the first
manual vetting process $\sim$35\% of targets required additional
scrutiny.  Most targets can be decided upon with the DV data products,
however some targets undergo scrutiny using additional data analysis
tools when tests using the standard data products are inconclusive or
unavailable.

\subsection{Flux Time Series Dispositioning}\label{sec:fluxvet}

The precision of the flux time series is sufficient to distinguish
some categories of stellar binaries that can mimic a transiting planet
\citep{BRO03,TOR04}.  In this section we describe three criteria
evaluated to determine whether the observed flux time series for a KOI
is consistent with a planet candidate or false positive.  The first
criterion investigates the presence of a ``secondary'' transit event
in the flux time series data with the same period as the KOI.  The
presence of a secondary suggests the signal originates from two
self-luminous bodies undergoing mutual eclipses and is a strong
indicator of a stellar binary false positive.  A phase-folded
median-filtered (filter window is the geometric mean of the transit
duration and orbital period) using an updated version of the PDC flux
time series \citep{STU12,SMI12} is visually examined for evidence of a
secondary.  Often a significant secondary also triggers an additional
TCE at the same period.  If an apparent secondary event is present,
the secondary is reexamined in both PA flux time series with a median
detrended filter applied and a PDC flux time series with a wavelet
whitening filter applied \citep[as employed during the planet search
  in TPS,][]{JEN02}.  A consistent and visually significant secondary
event in all filtering methods results in a false positive
disposition.  To prevent planetary candidates with secondary
occultations from being dispositioned as false positives, the
secondary event is fit for its depth to estimate the geometric albedo
and its uncertainty following \citet{ROW06}.  An estimated geometric
albedo ${\rm A_{g}}>$1.0 with 3 $\sigma$ confidence implies that a
statistically significant amount of flux is being emitted by the
planet beyond the expected amount due to reflection alone.  KOIs with
${\rm A_{g}}<$1.0 maintain their planet candidate status even with the
presence of a secondary since we cannot rule out a planet candidate
that produces a secondary through reflected light or thermal emission.

The second criterion measures whether there is a statistically
significant difference in the transit depth between alternating
events.  A statistically significant depth difference implies that the
KOI is a stellar binary with an orbital period twice the KOI ephemeris
period with similar primary and secondary depths (although different
enough to statistically confirm the depth difference).  The odd-even
depth test implements independent transit model \citep{MAN02} fits to
the odd- and even-numbered transit events and measures the statistical
difference in the resulting transit depths.  The results of the
transit model fits are available in the DV report summary.  A KOI is
designated a false positive if the odd-even depths are statistically
different ($>$3$\sigma$) from the model fits and a visual examination
of the phased odd-even light curves agrees with this assessment.  The
false positives identified by the odd-even depth test are confirmed
with model fits independent of the pipeline analysis using an
alternative filtering of the data that uses a median detrended PA flux
time series.

The final criterion examined in the flux times series is a qualitative
judgment as to the reliability and uniqueness of the transit signal.
The reliability of the transit signal is visually judged based upon
the several panels displaying the flux time series on the DV report
summary.  Having features of similar depth and duration in the phase
folded out-of-transit baseline data results in designating the KOI as
a false positive.  The process for examining the transit signal
reliability for KOIs is similar to the TCE triage steps (see
\S~\ref{sec:triage}).  However, the dispositioning process is more
thorough.  During KOI dispositioning an object can be subjected to
additional follow up analysis using data products beyond the DV report
summary when necessary.

\subsection{Pixel-level Time Series Dispositioning}\label{sec:centroidvet}

The second phase of dispositioning focuses on the pixel-level time
series data \citep{BAT13,BRY13} available at the Mikulski Archive for
Space Telescopes (MAST)\footnote{http://archive.stsci.edu}.  The
diagnostics based upon pixel-level data determine whether the
transit-like signal originates from the target or is spatially offset.
A transit signal that conclusively does not originate from the
designated target in the photometric aperture is designated a false
positive.  If the pixel-level data are consistent with a transit on
the target or the tests are inconclusive, then it is designated as a
planet candidate.  \citet{BRY13} describes the process for combining
these diagnostics into a decision for dispositioning the pixel-level
data.  We briefly describe the pixel-level diagnostics here.

There are two pixel diagnostics employed to locate the origin of the
transit signal.  The first diagnostic is a stellar image position time
series using a flux-weighted centroiding algorithm.  A statistically
significant offset in the flux-weighted centroid during transit
indicates that the target star is not isolated but has contaminating
flux in the aperture from other sources \citep{WIE96}.  The detector
row and column positions of the image are calculated using a
flux-weighted centroid algorithm for every cadence and converted into
a time series of RA and declination positions from a pixel to
coordinate transformation determined in PA.  The RA and declination
position time series is median-filtered with a window of 48
\kepler\ long cadences ($\sim$30 min in duration) and phase folded on
the KOI ephemeris.  In order to determine the change in image position
during transit, the \citet{MAN02} transit model using parameters from
the fit to the flux time series is fit to the centroid time series.
In this case, the only free parameter in the model is a scale factor
that converts the fixed transit model into the relative change in the
image position.  The statistical significance of the centroid offset
and its direction during transit is the diagnostic considered for
dispositioning.  Figures supporting the centroid time series fit are
provided in the DV products for each TCE \citep[see Figures~4~\&~5
  of][for a description of their use in dispositioning]{BRY13}.  Since
it indicates whether the stellar flux observed in the aperture is
composed of multiple sources, a significant change in the flux
weighted centroid during transit does not by itself indicate that the
source of the transit signal is not on the primary target of the
photometric aperture.  Thus, the flux weighted centroid information
plays a supporting role to the second pixel-level diagnostic described
next.

The second diagnostic considered from the pixel-level data is based on
consideration of flux difference images.  A flux difference image is
calculated for every observing quarter that contains transit events by
determining the average change in flux during transit based upon the
ephemeris of the TCE.  This is done on a pixel-by-pixel basis in order
to spatially locate the source of the transit signal on the detector.
A flux difference image is good quality and provides useful
information when it has the appearance of a stellar point spread
function (PSF) \citep[see Figure~11 of][for examples of a high quality
  flux difference image]{BRY13}.  Stellar crowding, low SNR transit
events, saturated stars, and various systematics can result in
difference images that are qualitatively inconsistent with a stellar
image and invalidate the results of the next pixel-level disposition
criterion.  In the test considered here, a visual inspection is
performed in order to determine whether a majority of the flux
difference images are consistent with a stellar PSF.  In the case when
a majority of the flux difference images are of poor quality, the KOI
is given a planet candidate disposition since the pixel-level
disposition criterion based upon the flux difference images will be
unreliable.

If the transit signal originates from the target of interest in the
photometric aperture, then the flux difference image will have the
appearance of a stellar point spread function (PSF) centered on the
target location.  A flux difference image that has the appearance of a
stellar PSF and has a statistically significant offset from the target
is evidence for a false positive.  To quantify the position of the
flux difference image, the \kepler\ Pixel Response Function (PRF)
model \citep[described in][and available through MAST]{BRY10} is fit
to the flux difference image.  A $\chi^{2}$ minimization solves for
the stellar position and brightness that minimizes the residual
between the observed and model images.  The out-of-transit position of
the target is determined by two methods: 1) adopting the KIC position
as the target position and 2) performing another PRF model fit to the
``direct'' image to determine the target position.  The direct image
is based upon averaging a contiguous, but limited, set of
out-of-transit images that occur before and after all transit events
in a quarter.  The set of contiguous out-of-transit images has the
same duration as the transit, but it is offset by 3 observing cadences
preceding and following the first and fourth contact of the transit event,
respectively.

The observed target position from the direct image is the preferred
comparison to the flux difference image, however, the KIC position
provides a more robust position when stellar crowding results in a
biased direct image position.  For each observing quarter an offset
and its uncertainty are calculated between the flux difference image
and both target position estimates.  The single-quarter results are
combined to derive a robust average offset across multiple quarters.
An average position offset $>$3$\sigma$ results in a false positive
designation.  In general, both the direct image position and KIC
position offsets should be $>$3$\sigma$ for a false positive
designation, however a single offset being significant is sufficient
for a false positive designation if a visual inspection of the target
scene warrants concluding any of the underlying assumptions of the
offset calculation are violated \citep[see][ for a thorough discussion
  of \kepler\ pixel-level data analysis]{BRY13}.

\subsection{Post Vetting Analysis}\label{sec:postvet}

The previous sections describe the criteria based upon \kepler\ data
that are investigated in order to designate a TCE as a planet
candidate or a false positive.  In this section, we describe
additional checks that are performed on the KOI sample.  During
analysis of the PRF fit results from the Q1-Q10 pipeline run used for
dispositioning, we identified an underestimate of the uncertainty in
the fitted position for small ($\lesssim 1.0\arcsec$) flux difference
image offsets \citep[see section 6.3 of][]{BRY13}.  To mitigate the
underestimate in the formal uncertainties, a systematic noise floor of
$\sigma=0.06\bar{6}$$\arcsec$ is added in quadrature to the reported
uncertainty in the angular offset between the flux difference image
and the direct image.  The KOIs impacted by this re-evaluation of the
centroid uncertainties were re-examined to provide more robust
centroid diagnostics.  The re-evaluation recognizes the systematic
noise floor in the centroid offsets by accepting offsets
$<$0.2$\arcsec$ as not significant independent of the formal
uncertainty, and transitions to accepting offsets $>$3$\sigma$ as
significant when the offset is $>$2.0$\arcsec$.  The re-evaluation of
the centroid offset significance in the transition range
0.2-2.0$\arcsec$, are evaluated as outlined in Section 6.3 of
\citet{BRY13}.  Approximately 25~centroid-based false positives from
the standard dispositioning process were designated as centroid-based
planet candidates from this re-evaluation.

It is possible for an astrophysical signal within the focal plane of
\kepler\ to contaminate the photometric aperture of an unrelated
target through several mechanisms: internal reflections, direct PRF
contamination, CCD saturation column bleed, video cross talk, and an
unexplained column anomaly mechanism \citep[][Coughlin et al., in
  preparation]{CAL10}.  Internal reflection and direct PRF
contamination can inject an additive low photon flux from a stellar
eclipsing binary (EB) into an unrelated target aperture so as to mimic
a transiting planet signal.  To identify this and other sources of
aperture contamination, we examine any matches between the KOI planet
candidate sample ephemerides to the ephemerides of known eclipsing
binaries from the \kepler\ Eclipsing Binary Catalog v3.0
\citep{PRS11,SLA11} and ground-based surveys (Coughlin et al., in
preparation).  KOIs with an ephemeris match to another KOI or EB are
visually examined to verify that phase folded flux time series of the
potential offending binary matches the KOI.  The KOI sample has a
substantial number of EB ephemeris matches, however most are already
dispositioned as false positive following dispositioning.  We changed
the dispositions of 5 KOIs from planet candidate to false positive due
to EB ephemeris matching.

In order to vet a KOI with \kepler\ pipeline data products, TPS needs
to identify a TCE that matches the ephemeris of the KOI.  Since each
instance of the pipeline is independent, there is no guarantee that a
matching TCE will be generated in a subsequent instance of the
pipeline.  Among the $\sim$1900 KOIs dispositioned in this study, 192
KOIs did not generate a TCE in the Q1-Q10 pipeline run.  For these 192
KOIs, subsequent pipeline runs (up to and including a Q1-Q12 pipeline
run) enabled 130 KOIs from this study to be dispositioned with
\kepler\ DV data products.  The remaining 62 KOIs were dispositioned
through manual analysis of the \kepler\ flux and pixel-level data
products.  Nearly half, 30, of the KOIs requiring manual analysis were
classified as false alarms since they no longer showed sufficient
significance in the flux time series to warrant a detection above the
MES$>$7.1$\sigma$ threshold, and were given a false positive
disposition.  The remaining 32 KOIs requiring manual analysis were
roughly evenly distributed among several possibilities: KOIs with one
or two transit events in the flux time series that don't generate a
TCE given the requirement of TPS for 3 transit events (these are given
a planet candidate disposition by default), KOIs with large transit
timing variations (since the \kepler\ pipeline assumes a constant
period in its search and DV analysis, these are given a planet
candidate disposition by default), and KOIs with deep transits that
were included on a list of targets not searched for planets (manual
analysis provided dispositions for these cases).

\subsection{Sample Completeness}\label{sec:complete}

The sample of planet candidates presented here represents an
inhomogeneous collection of the detections found with \kepler\ data.
This makes analysis of the underlying planet population from the
reported planet candidates challenging.  Ideally, the planet
candidates for planet population studies with \kepler\ would be
detected as TCEs and dispositioned using data products from the same
pipeline run; The Q1-Q8 \kepler\ planet candidate sample does not meet
this ideal.  The inhomogeneity arises due to having dispositioned a
subset of all KOIs using DV data products from the Q1-Q10 pipeline
run.  The two groups of KOI targets dispositioned in this study (the
new Q1-Q8 KOIs and KOI samples from \citet{BOR11A} and \citet{BOR11B})
included all the TCEs detected in the Q1-Q10 pipeline run.  However,
the KOI sample from \citet{BAT13} was not dispositioned in this study
and only includes KOIs detected in a Q1-Q6 (16 months) pipeline run.
Furthermore, targets that are not a part of the KOIs presented in this
study, but host planet candidates in the Q1-Q10 pipeline run are also
not represented in the current \kepler\ planet candidate sample.

\section{Stellar and Planetary Parameters}\label{sec:parameters}

In order to determine parameters of a planet, we fit a limb-darkened
transit signal model from \citet{MAN02} to the observed \kepler\ flux
time series.  The procedure is described in detail in \citet{BAT13},
though we briefly review it here.  The transit model is fit using a
scale-free set of variables (e.g., impact parameter ($b$), semi-major
axis to stellar radius ratio ($a/R_{\star}$), and planet to stellar
radius ratio, \Rplanet/\Rstar) that are weakly dependent (through the
limb-darkening coefficients) on the adopted stellar parameters.  The
fit is done assuming zero eccentricity and fixed limb-darkening
parameters according to the tables of \citet{CLA11}.  Providing an
accurate \Rplanet\ and $a$ from a transit light curve depends directly
on the stellar radius estimate as well as the orbit having zero
eccentricity in the case of $a$.  Before fitting the data using the
Levenberg-Marquardt $\chi^{2}$ minimization, the \kepler\ aperture
photometry output from the PA pipeline module is median detrended with
a 2~day window size in order to remove long time scale variability.
Since accuracy of the crowding metrics available for targets from the
KIC has not been studied in detail, no adjustment to the parameters
accounting for third-light dilution is applied.  The statistical
uncertainties on parameters are taken from the diagonal elements of
the covariance matrix.  The uncertainty on \Rplanet\ has the
uncertainty in \Rstar\ added in quadrature.

For some \kepler\ targets the combination of \Teff, \Logg, and
\Feh\ provided by the KIC is inconsistent with theoretical stellar
evolution calculations.  The stellar parameter combination adopted is
modified from the original KIC values to ensure consistency with the
Yale-Yonsei stellar isochrones \citep{YI01} following the procedure
described in \citet{BAT13}.  From an estimate of \Teff, \Logg, \Feh\,
a $\chi^{2}$ minimization is performed to determine the \Mstar\ and
\Rstar\ of the target.  For most targets with KIC photometry
available, the \Teff\ is adopted from \citet{PIN12} with a
corresponding average solar neighborhood metallicity of \Feh$=-0.2$.
For \Logg, the original KIC values are adopted.  A subset have the
input spectral parameters measured with high resolution spectroscopy
as part of the \kepler\ Follow-up Observation Program \citep{GAU10}
using the Spectroscopy Made Easy (SME) analysis package \citep{VAL96}.
Spectroscopic parameters also are available from the Stellar Parameter
Classification (SPC) analysis of \citet{BUC12} as well as \kepler\ based
asteroseismology results of \citet{HUB13}.  Stellar properties for
unclassified stars in the KIC are derived by interpolating typical
main-sequence colors and star properties given by \citet{SCH82} to the
observed 2MASS J-K colors \citep{SKR06}.  The stellar properties for
unclassified targets in the KIC have significant systematic
uncertainties and should be treated with caution.  These targets
hosting KOIs are indicated with the KIC unclassified column in
Table~\ref{tab:parm}.

\begin{figure}
\includegraphics[trim=0.3in 0.2in 0.55in 0.4in,clip=true,scale=0.52]{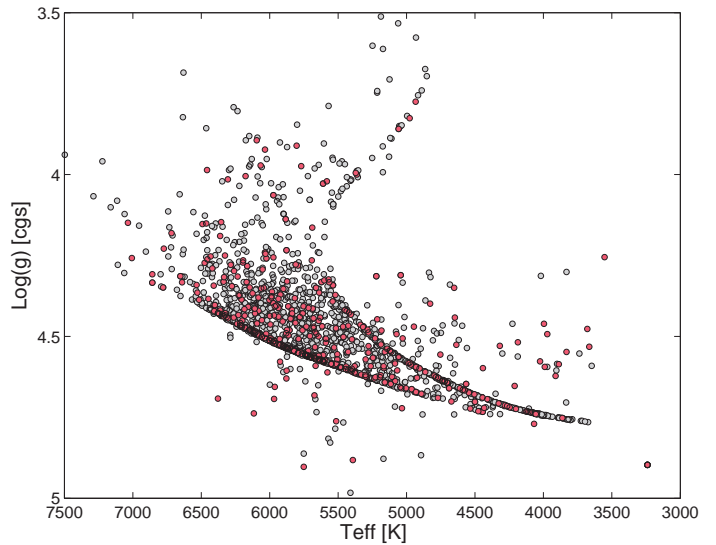}
\caption{
Stellar surface gravity as a function of stellar effective temperature estimates for the targets hosting new \kepler\ planet candidates (red points).  In addition, we show previously identified \kepler\ planet candidate hosts from \citet{BOR11A,BOR11B,BAT13} (gray points).
\label{fig:tefVlog}}
\end{figure}

Figure~\ref{fig:tefVlog} shows the \Teff\ versus \Logg\ stellar
parameters for the \kepler\ planet candidate sample hosts.  There is a
clear concentration of stellar properties on the outer limits of the
isochrones.  This results from forcing the calculated stellar
properties to match the closest isochrone available along with the
fact that most stars were assumed to have a fixed metallicity of
\Feh$=-0.2$.  The targets with stellar properties that scatter outside
the isochrone limits are targets for which no \Teff\ is available from
\citet{PIN12}, so the original KIC values were adopted (i.e.\ these
were not fitted to the isochrones).  The new Q1-Q8 \kepler\ planet
candidate hosts (red circles) follow the distribution of stellar hosts
to the previously identified planet candidates from
\citep{BOR11A,BOR11B,BAT13} (light gray circles), with a potential
deficit in the sub-giant and giant gravity regimes.

\section{Results}\label{sec:results}

Table~\ref{tab:parm} reports the properties of the KOI stellar host
and planet properties as described in \S~\ref{sec:parameters} and the
KOI planet candidate or false positive designation following the
procedure outlined in \S~\ref{sec:vetting}.  The table is
comprehensive for all KOIs known as of this Q1-Q8 study.
Table~\ref{tab:parm} reports the fitting of a limb-darkened transit
model to the \kepler\ flux data resulting in the most direct geometric
parameters: orbital period (\Porb), ephemeris epoch, transit model
depth at closest approach, transit duration (\Tdur), planet to star
radius ratio (\Rplanet/\Rstar), impact parameter ($b$), and semi-major
axis to \Rstar ratio ($a$/\Rstar).  The resulting SNR of the transit
model fit is also given in Table~\ref{tab:parm}.  Planet parameters
for KOIs with a transit fit SNR$<$10 have an increasing possibility
for systematic errors in the estimated parameters to become larger
than the provided uncertainty estimates.

Table~\ref{tab:parm} reports the stellar host property estimates:
stellar mass (\Mstar), stellar radius (\Rstar), effective temperature
(\Teff), and surface gravity (\Logg).  Combining the transit model
geometric parameters with the stellar parameters yields the indirect
planet properties: planet radius (\Rplanet) and planet equilibrium
temperature (\Teq) assuming a Bond albedo, $\alpha=0.3$, and full
surface redistribution of energy, $f=1.0$.  A more comprehensive table
with additional parameters and parameter uncertainties is available in
an interactive and searchable format from the NASA Exoplanet Archive
as the Q1-Q8 activity table.

The disposition column (Disp) of Table~\ref{tab:parm} reports the
status of a KOI as an integer where a value of 1 indicates a planet
candidate and 2 indicates a false positive disposition, respectively.
The newly dispositioned column (New KOI) of Table~\ref{tab:parm}
indicates a binary flag which has a value 1 when the KOI was
dispositioned with Q1-Q10 or more data products following the
procedure outlined in this study (see \S~\ref{sec:vetting}) and a
value 0 when the KOI was excluded from dispositioning because the
disposition status is adopted from \citet{BAT13}, \citet{BRY13}, or
the KOI is a confirmed \kepler\ planet.  The indeterminate period
column in Table~\ref{tab:parm} indicates KOIs for which there is not a
reliable period available since the flux time series only contains a
single transit/eclipse event or the multiple events that are present
do not enable uniquely assigning events to one or several candidates.
The KOIs with a 1 in the indeterminate period column of
Table~\ref{tab:parm} are subsequently not included in any statistical
counts or figures for the rest of this paper as the information for
these KOIs has considerable uncertainty.

\begin{figure}
\includegraphics[trim=0.3in 0.2in 0.55in 0.4in,clip=true,scale=0.52]{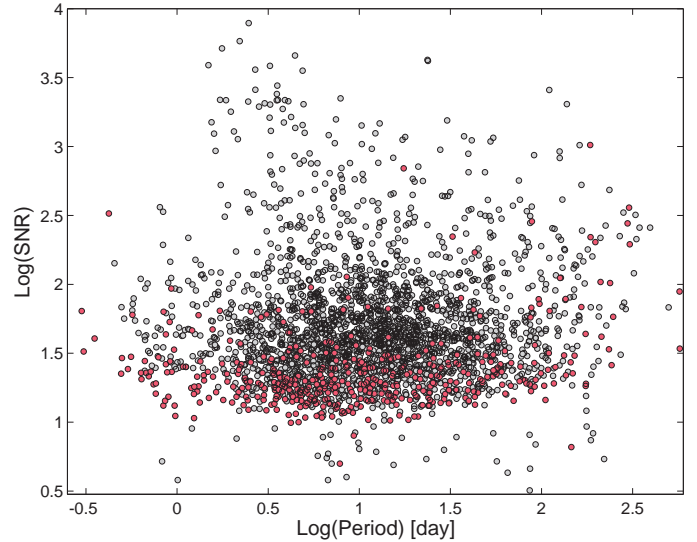}
\caption{
The transit SNR using Q1-Q10 data as a function of orbital period for the new \kepler\ planet candidates (red points). In addition, we show previously identified \kepler\ planet candidates from \citet{BOR11A,BOR11B,BAT13} (gray points).
\label{fig:perVsnr}}
\end{figure}

\begin{figure}
\includegraphics[trim=0.3in 0.2in 0.6in 0.4in,scale=0.5,clip=true]{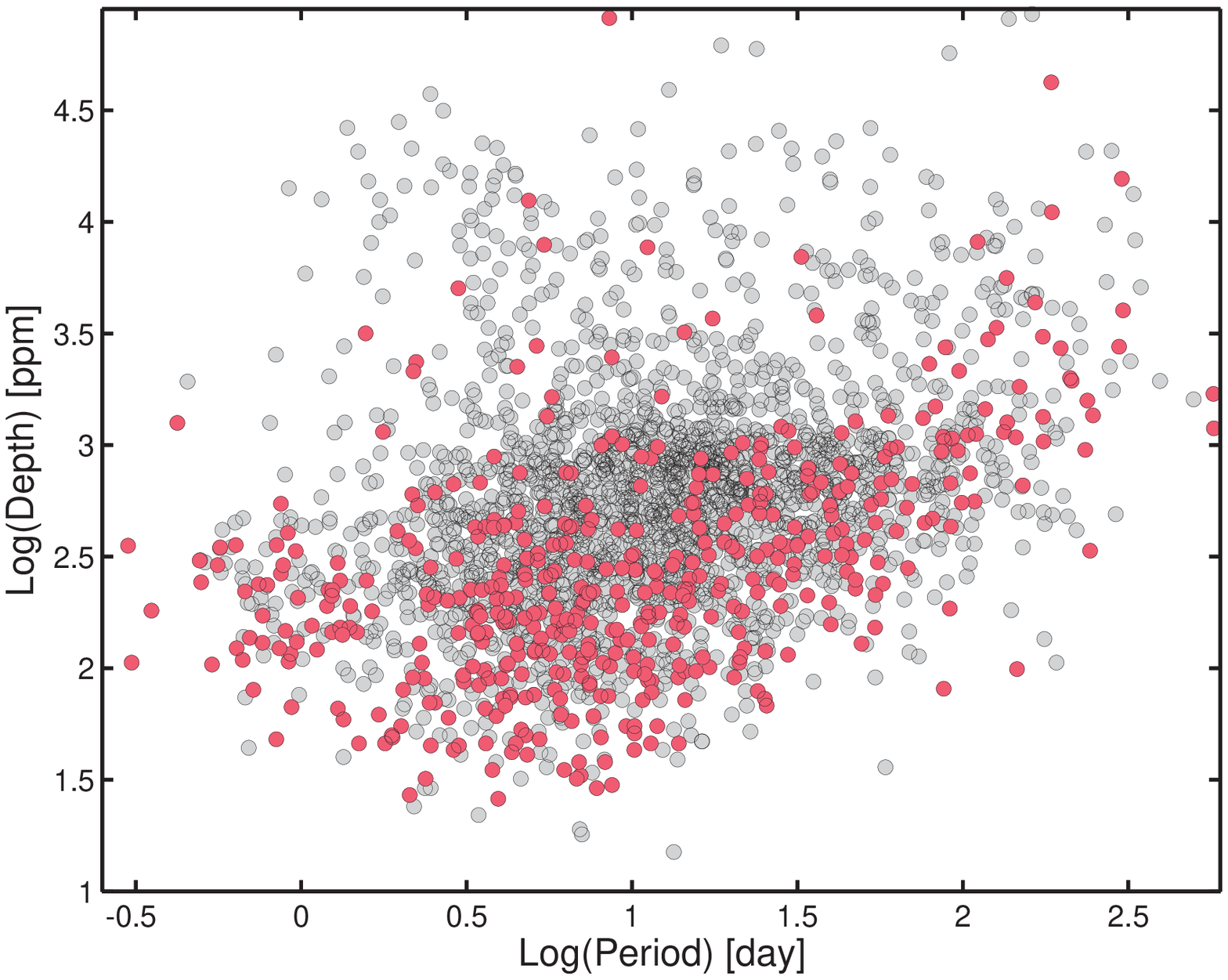}
\includegraphics[trim=0.3in 0.2in 0.6in 0.4in,scale=0.5,clip=true]{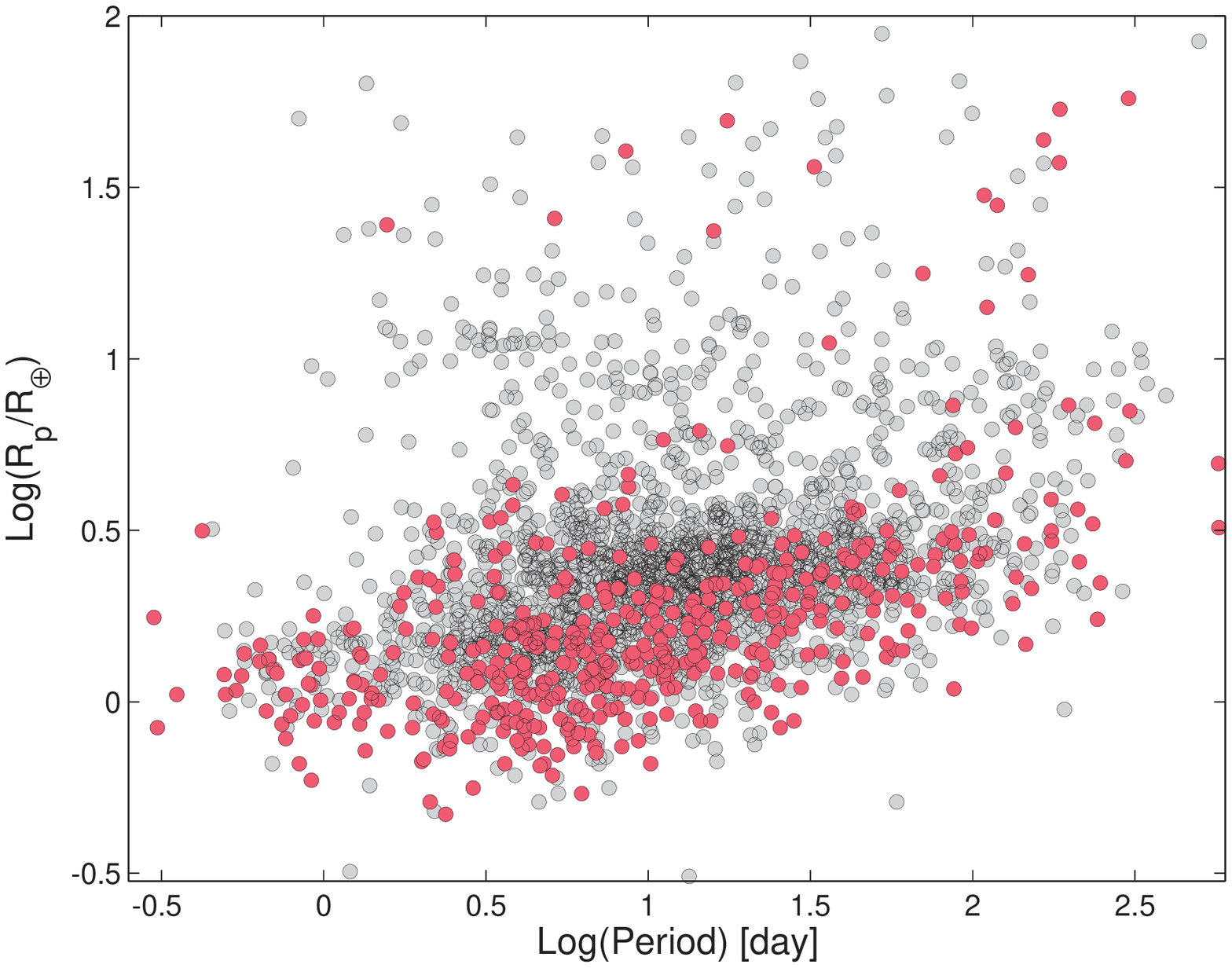}
\caption{
{\it Left:} The transit depth as a function of orbital period for the new \kepler\ planet candidates (red points). In addition, we show previously identified \kepler\ planet candidates from \citet{BOR11A,BOR11B,BAT13} (gray points). {\it Right:} Resulting \kepler\ planet candidate radii relative to \Rear\ from limb-darkened transit model fits for the new \kepler\ planet candidates (red points) and previously published \kepler\ planet candidates (gray points).
\label{fig:perVdep}}
\end{figure}

\begin{figure}
\includegraphics[trim=0.3in 0.2in 0.4in 0.0in,clip=true,scale=0.5]{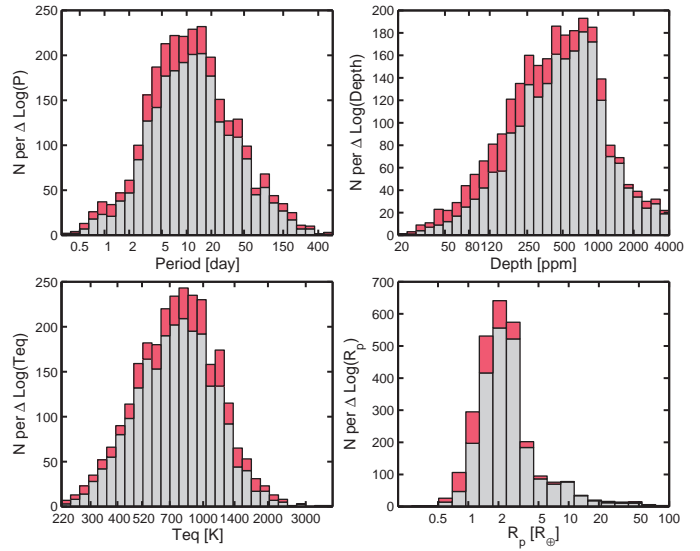}
\caption{
Distribution of \kepler\ planet candidate parameters: orbital period ({\it upper left}), transit depth ({\it upper right}), planet equilibrium temperature ({\it lower left}), and planet radius ({\it lower right}) using equal logarithmically spaced intervals.  The contribution for new \kepler\ planet candidates (red) in each interval are shown on top of the previously published \kepler\ planet candidates (gray).
\label{fig:bar}}
\end{figure}

\begin{figure}
\includegraphics[trim=0.3in 0.2in 0.4in 0.0in,clip=false,scale=0.5]{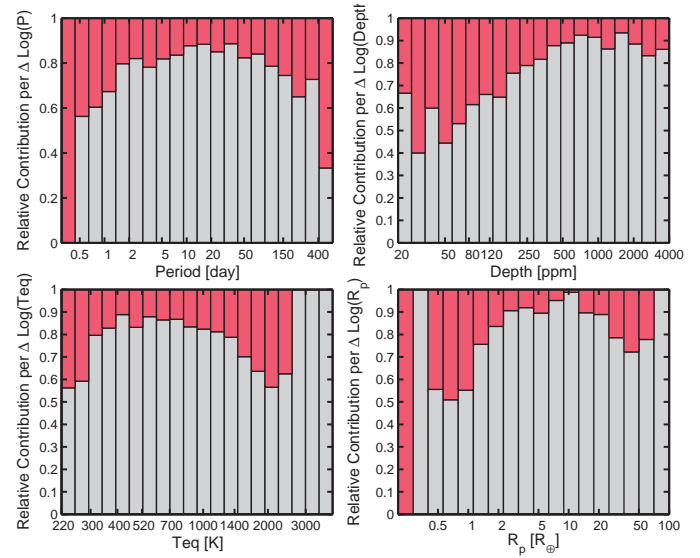}
\caption{
Normalized contribution of new \kepler\ planet candidates (red) relative to the previously published \kepler\ planet candidates (gray) over equal logarithmically space intervals for several planet parameters: orbital period ({\it upper left}), transit depth ({\it upper right}), planet equilibrium temperature ({\it lower left}), and planet radius ({\it lower right}).  This figure complements what is shown in Figure~\ref{fig:bar}.
\label{fig:barfrac}}
\end{figure}

\begin{figure}
\includegraphics[trim=0.3in 0.2in 0.6in 0.4in,scale=0.5,clip=true]{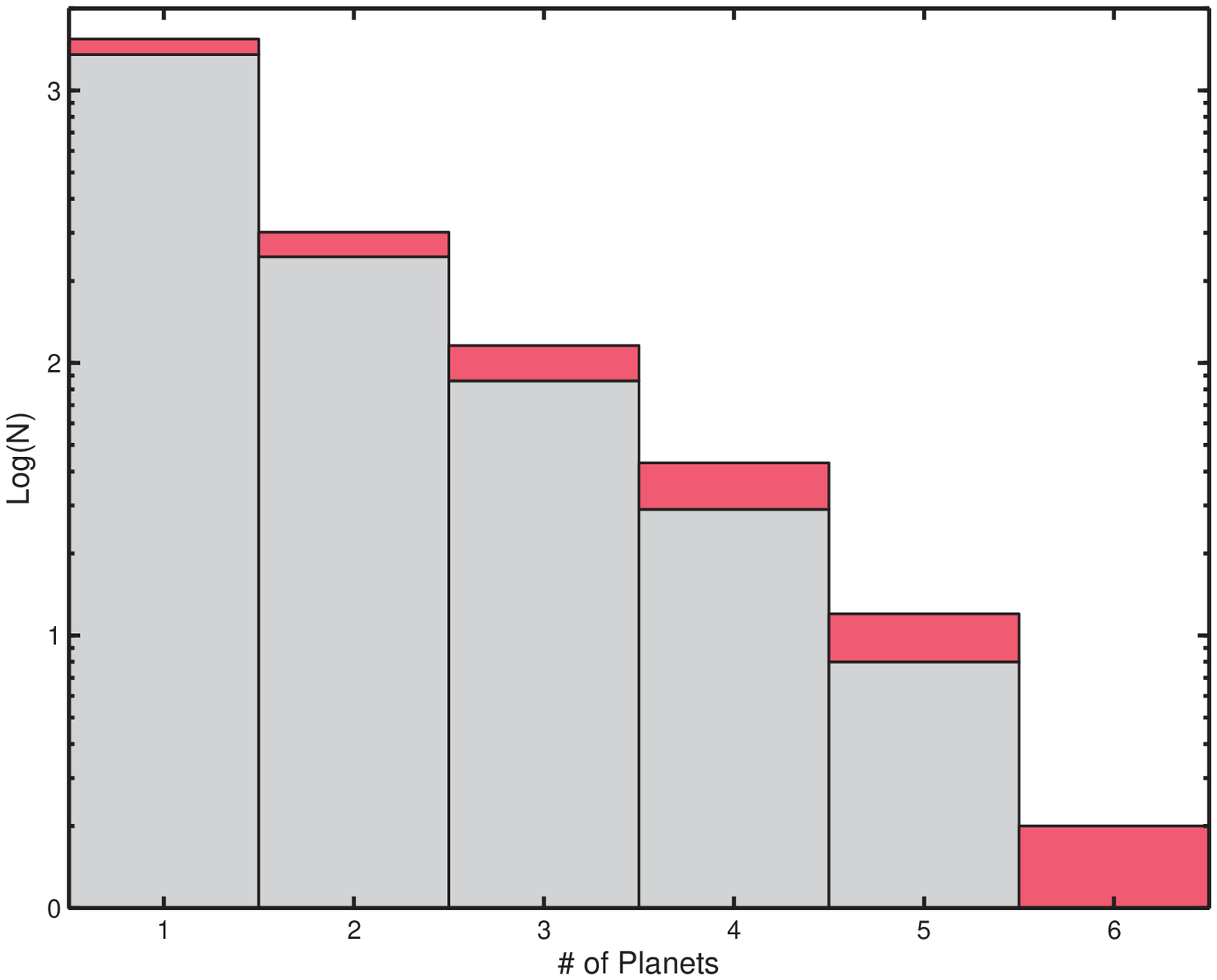}
\includegraphics[trim=0.3in 0.2in 0.6in 0.4in,scale=0.5,clip=true]{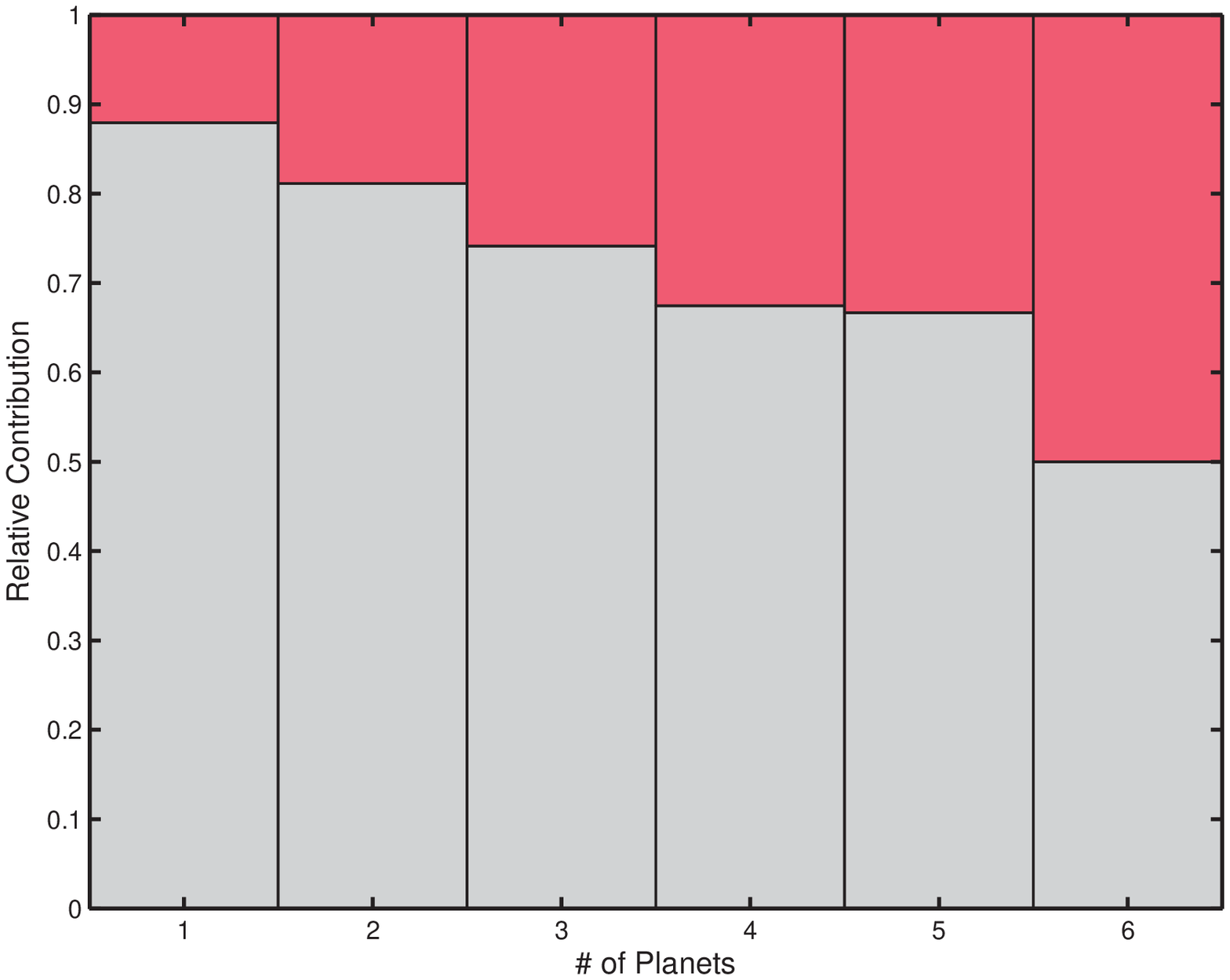}
\caption{
{\it Left:} Logarithm of the number of \kepler\ planet candidate hosts having planet multiplicity as indicated on the abscissa. The contribution for new \kepler\ planet candidates (red) in each planet multiplicity are shown on top of the previously published \kepler\ planet candidates (gray).  {\it Right:}  Normalized contribution of new \kepler\ planet candidate hosts (red) relative to the previously published \kepler\ planet candidate hosts (gray) for each planet multiplicity.
\label{fig:multibar}}
\end{figure}

\begin{figure}
\includegraphics[trim=0.3in 0.2in 0.4in 0.0in,clip=false,scale=0.5]{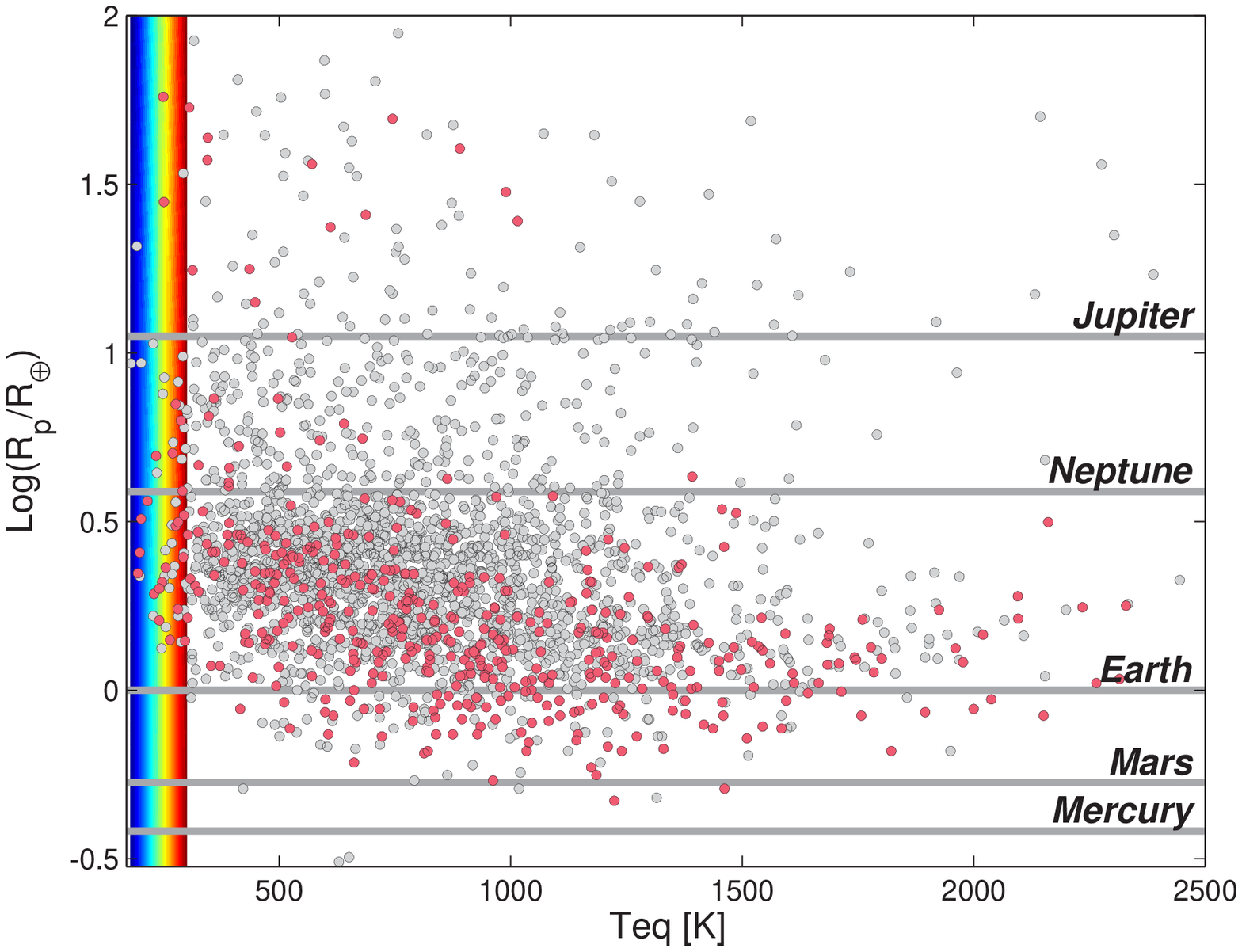}
\caption{
The planet radius relative to \Rear\ as a function of the planet equilibrium temperature (assuming $\alpha=0.3$ and $f=1.0$) for the new \kepler\ planet candidates (red points) and the previously identified \kepler\ planet candidates from \citet{BOR11A,BOR11B,BAT13} (gray points).  Representative Solar System planet radii are labeled (solid horizontal lines), and the illustrative range of the HZ (color shaded vertical band) adopted in this study is shown.
\label{fig:teqVrpl}}
\end{figure}

Figures~\ref{fig:perVsnr}-\ref{fig:teqVrplzoom} illustrate the
properties of the \kepler\ planet candidate sample that is tabulated
in Table~\ref{tab:parm} with the disposition flag, Disp=1.  The new
\kepler\ planet candidates identified in this study are indicated in
the figures using red markers and the \kepler\ planet candidates
identified in the earlier studies of \citet{BOR11A,BOR11B,BAT13} using
light gray markers.  Figure~\ref{fig:perVsnr} shows the transit SNR
(measured after detrending of the flux time series) resulting from the
Q1-Q10 data limb-darkened transit model fits as a function of orbital
period.  The new \kepler\ planet candidates (red circles) are
concentrated at low SNR relative to the previously identified
\kepler\ planet candidates (gray circles).  The rough floor in SNR is
slightly elevated from the 7.1 threshold level, since the original
detection being from Q1-Q8 data whereas the SNR is calculated from
Q1-Q10 data.  The population of high SNR planet candidates that are
new discoveries arose for several reasons.  At long periods, high SNR
signals, even from a single transit event, previously did not have the
requisite three transit events for detection in the pipeline.  At
short periods, high SNR detections occur for targets that were added
to the observing sample in later observing quarters (typically as part
of the \kepler\ Guest Observing
program\footnote{http://keplerscience.arc.nasa.gov}).  The criteria
(such as transit depth or obvious eclipsing binary signatures) for
cataloging KOIs has varied through time.  The pipeline software
continues to develop increasing sensitivity.

The left panel in Figure~\ref{fig:perVdep} shows the depth of transit
(as measured from the best-fit transit model minimum relative flux) as
a function of the orbital period.  The new \kepler\ planet candidates
identified in this study (red circles) populate a region of parameter
space indicative of them being lower SNR detections (i.e.\ toward
shallower depth and longer orbital period) than the previously
identified \kepler\ planet candidates (light gray circles).
Empirically, the sensitivity to planet candidates with a similar depth
to 1 \Rear\ transiting in front of a 1 \Rsun\ host ($\Delta$=84 ppm)
drops off considerably beyond a $\sim$30 day orbital period using
Q1-Q8 \kepler\ data.  The right panel in Figure~\ref{fig:perVdep}
shows the \kepler\ planet candidate radii estimates (as measured from
the best-fit transit model and stellar parameters described in
\S~\ref{sec:parameters}) as a function of orbital period.  Due to the
presence of host stars of later-type than the Sun, the sensitivity to
1 \Rear\ planets extends to a longer orbital period and drops off
strongly beyond a $\sim$55 day orbital period using Q1-Q8
\kepler\ data.

After re-evaluating the \kepler\ planet candidates from
\citet{BOR11A,BOR11B}, adopting the results of the \kepler\ planet
sample from \citet{BAT13}, and the \NewPC\ new \kepler\ planet
candidates introduced in this study, there are \TotPC\ \kepler\ planet
candidates in total.  Thus, the new planet candidates increase the
sample size by 21\%.  Figure~\ref{fig:bar} shows the distribution of
the \kepler\ planet candidate sample in equal logarithmically spaced
bins for the most interesting planet properties: orbital period,
transit depth, radius, and equilibrium temperature, starting in the
upper left panel and continuing clockwise, respectively.  In each bin
the red bar (on top) represents the count from the new planet
candidates and the light gray bar (on bottom) represents the count
from the previous planet candidate sample.  The observed distribution
of \kepler\ planet candidates represents the underlying planet
distribution as shaped by the sensitivity of the \kepler\ instrument,
pipeline planet search, and planet candidate evaluation process.  We
defer deriving the underlying planet population from \kepler\ planet
candidate samples to future work (see \S~\ref{sec:complete}).

Due to their concentration toward low SNR (see
Figure~\ref{fig:perVsnr}), the new planet candidates preferentially
make a larger contribution to the small planet and long period
distributions shown in Figure~\ref{fig:bar}.  In order to elucidate
the contribution of the new planet candidates in parameter space, we
complement the planet candidate distribution of Figure~\ref{fig:bar}
by showing in Figure~\ref{fig:barfrac} the relative contribution of
the new planet candidates and old planet candidates.  For instance, of
all \kepler\ planet candidates $\sim$40\% with
\Rplanet$\sim$1\Rear\ (lower right panel) are new in this study.  For
the cool \Teq$<$300~K \kepler\ planet candidates $\sim$40\% are new in
this study.

Multiple planet systems continue to be an important contribution to
the \kepler\ planet candidate sample \citep{FOR11,LIS11B,FAB12B}.  The
\TotPC\ \kepler\ planet candidates are distributed amongst
\TotTargets\ stellar hosts.  Of the planet candidate stellar hosts,
475 (23\%) host multiple observed transiting planet candidates.
Despite the multiple planet hosts being in the minority, their high
observed multiplicity (up to 6) results in 1,196 (46\%) of the planet
candidates residing in multiple systems.  The systems with 6 planet
candidates are the confirmed system Kepler-11b-g \citep[KOI
  157][]{LIS11A} and KOI 351.01-0.06.  The left panel of
Figure~\ref{fig:multibar} shows the logarithm of the number of stellar
hosts with planet candidates as a function of planet candidate
multiplicity.  The relative contribution of the new planet candidates
to each multiplicity bin is shown in the right panel of
Figure~\ref{fig:multibar}.  The new planet candidates contribute a
higher fraction of the rare, high multiplicity systems.

Figure~\ref{fig:teqVrpl} shows the planet candidate radius as a function
of equilibrium temperature assuming an Earth-like Bond albedo,
$\alpha=0.3$, and redistribution of energy over the full surface.
Horizontal lines indicate radii of the Solar System planets for
reference, and the shaded vertical band indicates the adopted
habitable zone (HZ) region for the possible existence of liquid water.
Specifying the boundary of the HZ to support anthropocentrist life is
an active area of research \citep{SEL07,KAS11,KOP13,ZSO13}.  In this
study we adopt 180$\le$\Teq$\le$300 K as the HZ boundary based
upon the guidance of \citet{KAS11}.  The adoption of a sharp boundary
for the HZ oversimplifies the complicated effects that can influence
the ability for a planet to maintain a reservoir of liquid water (bulk
planet composition, atmospheric composition, cloud and surface
dynamics, etc.), and oversimplifies the significant observational
uncertainty in equilibrium temperature for determining whether a
planet lies in the HZ.  However, since we only have the most basic
information available about the planet candidate orbital period and
stellar host properties the simplified HZ adopted here is sufficient
for our purposes of providing descriptive statistics of the \kepler\
planet candidate sample.

Figure~\ref{fig:teqVrplzoom} also shows the planet candidate radii as
a function of equilibrium temperature, but details the narrow HZ
region.  The point size is proportional to the orbital period of the
planet candidate.  The largest point size corresponds to P$>$300 day
and linearly decreases to the smallest point size corresponding to
P$<$40 day.  For fixed equilibrium temperature a larger point size
indicates the planet candidate has a stellar host with an earlier
spectral type.  The vertical dashed line shows the \Teq=255~K for
Earth.  Star symbols indicate the Solar System planets Mars, Earth, \&
Venus from left to right, respectively.  Overall, there are 57 HZ
\kepler\ planet candidates of which 23 (40\%) are new in this study.  By
adopting a more restrictive \Teq$<$270~K upper limit boundary to the
HZ \citep{SEL07}, there are 32~HZ \kepler\ planet candidates of which 14
(44\%) are new in this study.

From the set of new \kepler\ planet candidates in the HZ the three KOIs
closest in radius and equilibrium temperature to Earth are KOI 172.02,
3010.01, and 1422.04.  Subsequent to the first release of the new KOIs
from this study at the NASA Exoplanet
Archive\footnote{http://exoplanetarchive.ipac.caltech.edu/docs/tce\_releasenotes\_q1q12.pdf},
KOI~172.02 was confirmed as Kepler-69c \citep{BAR13}.  Of these three
HZ KOIs new in this study, Kepler-69c has the longest orbital period,
P=242~day, and orbits a Solar-type G4V host.  The more detailed
analysis of the Kepler-69 system \citep{BAR13} yields a similar
\Rplanet=1.7\Rear\ (versus \Rplanet=1.74\Rear) and warmer \Teq=299~K
(versus \Teq=281~K) planet than adopted in this study.  KOI 3010.01
and 1422.04 orbit significantly cooler \Teff$\sim$ 4000~K, and later
(\Rstar$\sim$0.5~\Rsun) spectral type hosts, and both have orbital
periods $\sim$60~day.  The planet candidate KOI 3010.01 has an
estimated \Rplanet=1.4~\Rear\ and \Teq=264~K. The planet candidate KOI
1422.04 has an estimated \Rplanet=1.6~\Rear\ and \Teq=241~K.

\section{Conclusion}\label{sec:conclusion}

This study examines the potential planet candidate signals generated
by the \kepler\ pipeline software searching Q1-Q8 ($\sim$ 2 Years) of
\kepler\ data.  The Q1-Q8 search resulted in $\sim$13,400 targets with
potential planet candidate signals, which was reduced to $\sim$480 new
viable targets for planet candidate signals from an initial
examination of the pipeline generated diagnostics (see
\S~\ref{sec:data}).  The viable planet candidates are assigned KOI
numbers and undergo additional scrutiny in order to classify them into
the \kepler\ planet candidate and false positive categories (see
\S~\ref{sec:vetting}).

In addition to the new Q1-Q8 KOIs, in this study we re-examined KOIs
from the first two \kepler\ planet candidate samples
\citep{BOR11A,BOR11B} in order to take advantage of the substantially
increased data baseline and a more uniform set of dispositioning
criteria and procedures.  Overall, between the new Q1-Q8 KOIs and
re-evaluation of the \citet{BOR11A,BOR11B} KOIs, we classified
$\sim$1,900 KOIs.  To classify these two groups of KOIs we took
advantage of improved pipeline data products and software using Q1-Q10
data.  The total \kepler\ planet candidate sample reported in this
study combines the new Q1-Q8 KOIs, the \citet{BOR11A,BOR11B} KOI
sample re-evaluation, and the KOI sample and classification of
\citet{BAT13}, which is adopted without further scrutiny.  Since the
KOIs of \citet{BAT13} (evaluated with Q1-Q6 \kepler\ data) were not
re-evaluated uniformly with the other KOIs (evaluated with Q1-Q10
\kepler\ data), the total \kepler\ planet candidate sample represents
an inhomogeneous collection of the detections available with
\kepler\ data making analysis of the underlying planet population from
the reported planet candidates challenging (see
\S~\ref{sec:complete}).

The current \kepler\ planet candidate sample provides researchers with a
well vetted sample of individual planet candidates and new and
expanded multiple planet candidate systems worthy of follow-up
observations and scientific study.  The total \kepler\ planet candidate
sample count from this study is \TotPC, with \NewPC\ (21\%) 
new with the Q1-Q8 data analysis.  The planet candidate gains are
concentrated at lower SNR than the previous samples (see
Figure~\ref{fig:perVsnr}); the new Q1-Q8 planet candidates
contribute significantly ($\sim$40\%) to the population of the \kepler\
planet candidates having \Rplanet$\sim$1\Rear\ and to the population
in the HZ of their stellar hosts.  Multiple planet systems continue to
be an important contribution to the \kepler\ planet candidate sample
\citep{FOR11,LIS11B,FAB12B}.  The \TotPC\ \kepler\ planet candidates are
distributed amongst \TotTargets\ stellar hosts.  Despite the multiple
planet hosts being in the minority (23\% of all stellar hosts), their
high observed multiplicity (up to 6) results in 46\% of the planet
candidates residing in multiple systems.

The Q1-Q8 \kepler\ planet candidate sample provides a rich population of
objects which help elucidate the planet formation, planetary
dynamics, stellar properties, and planet population statistics that
govern the existence of planets in The Milky Way.  We look forward to
the expanded \kepler\ discoveries that follow from the \kepler\ planet
candidates studied here along with the future discoveries enabled by
the larger baseline (Q1-Q17) of the recently completed \kepler\ 3- and
4-wheel modes of operation.

\acknowledgments

Funding for this Discovery mission is provided by NASA's Science
Mission Directorate.  MH and DH are supported by an appointment to the NASA
Postdoctoral Program at Ames Research Center, administered by Oak
Ridge Associated Universities through a contract with NASA.

\appendix
\section{Additional KOI Vetting}

This study specifically did not examine KOIs that were introduced in
the KOI sample of \citet{BAT13}.  However, 220 KOIs were given false
positive designations during the vetting of the planet candidate
sample from \citet{BAT13}, but these false positives were not
published or documented.  Because false positive identification
techniques have improved since \citet{BAT13}, we re-examined these 220
KOIs using the methods described in this paper.  The data
products used in the dispositioning were from a variety of pipeline
runs (using $\geq$Q1-Q10), and some of the KOIs required manual
analysis of \kepler\ data.  The column `Appendix KOI' in
Table~\ref{tab:parm} is a binary flag and indicates whether the KOI
belongs to the group of 220 previously unpublished KOIs with an
initial assessment as false positives.  As indicated in the column
`Disp' in Table~\ref{tab:parm}, 27 of these 220 KOIs have been
restored to planet candidate status following their re-examination.

\begin{figure}
\includegraphics[trim=0.3in 0.2in 0.4in 0.0in,clip=false]{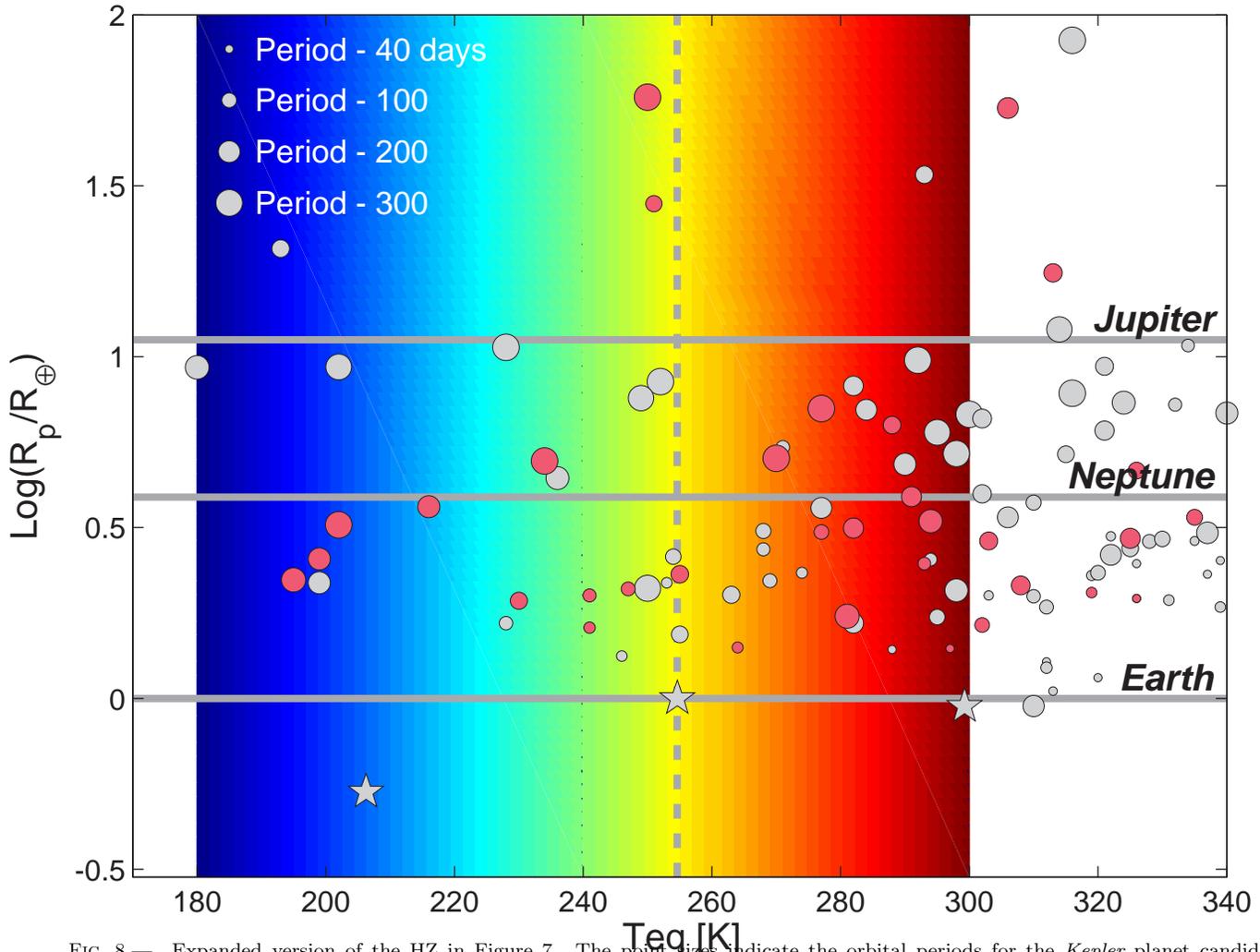}
\caption{
Expanded version of the HZ in Figure~\ref{fig:teqVrpl}.  The point sizes indicate the orbital periods for the \kepler\ planet candidates, highlighting the longest orbital period planet candidates in the HZ with a larger point size.  The largest point size is for candidates with $P>300$ day and linearly decrease in size down to $P<40$ day for the smallest point size.  This figure details the illustrative range of the HZ (color shaded vertical band) adopted in this study.  Representative Solar System planet radii (solid horizontal lines) and Earth's \Teq=255 K (dashed vertical line) are shown.  Star symbols indicate the Solar System planets Mars, Earth, \& Venus from left to right, respectively.
\label{fig:teqVrplzoom}}
\end{figure}

\begin{deluxetable}{@{\hspace{2 pt}}c@{\hspace{2 pt}}c@{\hspace{2 pt}}c@{\hspace{2 pt}}c@{\hspace{2 pt}}c@{\hspace{2 pt}}c@{\hspace{2 pt}}c@{\hspace{2 pt}}c@{\hspace{2 pt}}c@{\hspace{2 pt}}c@{\hspace{2 pt}}c@{\hspace{2 pt}}c@{\hspace{2 pt}}c@{\hspace{2 pt}}c@{\hspace{2 pt}}c@{\hspace{2 pt}}c@{\hspace{2 pt}}c@{\hspace{4 pt}}c@{\hspace{4 pt}}c@{\hspace{2 pt}}c@{\hspace{2 pt}}c@{\hspace{2 pt}}c@{\hspace{2 pt}}}
\tabletypesize{\scriptsize}
\tablewidth{0pt}
\tablecaption{{\rm KOI Ephemeris, Planet Parameters, Stellar Parameters, \& Disposition Status}\tablenotemark{a}}
\startdata
\hline
\hline
KOI & Kepler Id & \Porb\ & Epoch\tablenotemark{b} & Depth & a/\Rstar\ & b & \Tdur\ & \Rplanet/\Rstar\ & SNR & \Rstar & \Mstar & \Teff & \Logg\ & \Rplanet\ & Teq & Disp\tablenotemark{c} & New\tablenotemark{d} & Appendix\tablenotemark{e} & Disp\tablenotemark{f} & Indeterminate\tablenotemark{g} & KIC\tablenotemark{h}  \\
 &  & \brk{day} &  & \brk{ppm} &  &  & \brk{hr} & & & \brk{\Rsun} & \brk{\Msun} & \brk{K} & \brk{cgs} & \brk{\Rear} & \brk{K} & & KOI & KOI & This Study & Period & unclassified  \\
\hline
   1.01 &  11446443 &     2.470613 &  55.762566 &  14284.0 &   8.445 & 0.8222 &  1.725 & 0.12437 &  7856.10 &  1.06 &  0.99 &  5814.0 & 4.381 &  14.45 & 1394.0 & 1 & 0 & 0 & 0 & 0 & 0 \\
   2.01 &  10666592 &     2.204735 &  54.357802 &   6713.0 &   4.681 & 0.1282 &  3.877 & 0.07545 &  5812.40 &  2.71 &  1.66 &  6264.0 & 3.792 &  22.32 & 2303.0 & 1 & 0 & 0 & 0 & 0 & 0 \\
   3.01 &  10748390 &     4.887800 &  57.812537 &   4323.0 &  16.681 & 0.0286 &  2.368 & 0.05770 &  1433.40 &  0.74 &  0.79 &  4766.0 & 4.592 &   4.68 &  794.0 & 1 & 0 & 0 & 0 & 0 & 0 \\
   4.01 &   3861595 &     3.849372 &  90.525922 &   1340.0 &   4.481 & 0.9462 &  2.928 & 0.04152 &   284.00 &  2.60 &  1.61 &  6391.0 & 3.812 &  11.81 & 1923.0 & 2 & 0 & 0 & 1 & 0 & 0 \\
   5.01 &   8554498 &     4.780329 &  65.973245 &    966.0 &   7.560 & 0.9507 &  2.012 & 0.03651 &   456.20 &  1.42 &  1.15 &  5861.0 & 4.193 &   5.66 & 1279.0 & 1 & 0 & 0 & 1 & 0 & 0 \\
 ... & & & & & & & & & & & & & & & & & & & & & \\
3145.02 &   1717722 &     0.977314 &  64.698560 &    207.0 &   5.032 & 0.0307 &  1.262 & 0.01461 &    11.10 &  0.63 &  0.72 &  4771.0 & 4.690 &   1.01 & 1284.0 & 1 & 1 & 0 & 1 & 0 & 0 \\
3146.01 &  10908248 &    39.859247 &  87.772400 &    157.0 &  36.775 & 0.0261 &  8.355 & 0.01110 &    14.30 &  1.08 &  0.97 &  5900.0 & 4.353 &   1.31 &  570.0 & 1 & 1 & 0 & 1 & 0 & 0 \\
3147.01 &   7534267 &    39.441307 & 102.146160 &    197.0 &  42.945 & 0.0416 &  7.098 & 0.01246 &    17.00 &  0.86 &  0.97 &  5765.0 & 4.554 &   1.17 &  499.0 & 1 & 1 & 0 & 1 & 0 & 0 \\
3148.01 &  10611420 &    11.505399 &  68.098080 &    288.0 &  26.146 & 0.6919 &  2.500 & 0.01657 &    14.30 &  1.07 &  1.15 &  6354.0 & 4.438 &   1.94 &  896.0 & 2 & 1 & 0 & 1 & 0 & 0 \\
3149.01 &  10196493 &     9.811017 &  67.186760 &     61.0 &  16.959 & 0.8950 &  2.086 & 0.00868 &     4.80 &  1.00 &  0.94 &  5799.0 & 4.406 &   0.95 &  864.0 & 2 & 1 & 0 & 1 & 0 & 0 \\
\hline
\enddata
\label{tab:parm}
\tablenotetext{a}{A version of the Table with comprehensive reporting of parameters and their uncertainties is published in the Q1-Q8 KOI activity table at the NASA Exoplanet Archive http://exoplanetarchive.ipac.caltech.edu}
\tablenotetext{b}{BJD=Epoch+2454900.0}
\tablenotetext{c}{Disposition ; 1 = Planet Candidate ; 2 = False Positive}
\tablenotetext{d}{New KOI to \kepler\ Q1-Q8 sample ; 1 = True ; 0 = False}
\tablenotetext{e}{KOI from Appendix A. ; 1 = True ; 0 = False}
\tablenotetext{f}{Disposition for KOI was updated during this study ; 1 = True ; 0 = False}
\tablenotetext{g}{Single transit or ambiguous period KOI ; 1 = True ; 0 = False}
\tablenotetext{h}{Stellar parameters unclassified in the \kepler\ Input Catalog ; 1 = True ; 0 = False}
\tablecomments{Table published in its entirety in the journal electronic addition.  A portion is displayed here to illustrate its form and content.}
\end{deluxetable}

\end{document}